\documentclass[11pt]{article}
\pdfoutput=1
\usepackage{jheppub}
\usepackage{graphicx}
\usepackage{amssymb}
\usepackage{subcaption}
\usepackage{amsmath}
\usepackage{slashed}
\usepackage{hyperref}
\usepackage{caption}
\usepackage{xcolor}
\usepackage{dsfont}

\newcommand\eq[1]{eq.~(\ref{eq:#1})}

\newcommand{\sn}[1]{\S~\ref{sec:#1}}

\newcommand\mc[1]{{\mathcal{#1}}}

\newcommand\CA{{\mc{A}}}

\newcommand\CD{{\mc{D}}}

\newcommand\CM{{\mc{M}}}

\newcommand\CT{{\mc{T}}}
\newcommand\CZ{{\mc{Z}}}

\newcommand\rar{\rightarrow}
\newcommand\Dslash{D\!\!\!\!\slash}
\newcommand\eps{\epsilon}

\newcommand\vphi{\varphi}

\newcommand\bpsi{\bar{\psi}}
\newcommand\Rfull{\mathbb{R}^{2,1}}
\newcommand\Rplus{\mathbb{R}^{2,1}_+}

\def\shrug{\texttt{\raisebox{0.75em}{\char`\_}\char`\\\char`\_\kern-0.5ex(\kern-0.25ex\raisebox{0.25ex}{\rotatebox{45}{\raisebox{-.75ex}"\kern-1.5ex\rotatebox{-90})}}\kern-0.5ex)\kern-0.5ex\char`\_/\raisebox{0.75em}{\char`\_}}}

\title{3d Abelian Dualities with Boundaries}

\preprint{\today}

\author[a]{Kyle Aitken,}
\author[a]{Andrew Baumgartner,}
\author[a]{ Andreas Karch,}
\author[a,b]{and Brandon Robinson}
\affiliation[a]{Department of Physics, University of Washington, Seattle, Wa, 98195-1560, USA}
\affiliation[b]{School of Physics and Astronomy and STAG Research Centre,
University of Southampton, Highfield, SO17 1BJ}
\emailAdd{kaitken@uw.edu}
\emailAdd{baum4157@uw.edu}
\emailAdd{akarch@uw.edu}
\emailAdd{b.j.robinson@soton.ac.uk}

\abstract{We establish the action of three-dimensional bosonization and particle-vortex duality in the presence of a boundary, which supports a non-anomalous two-dimensional theory. We confirm our prescription using a microscopic realization of the duality in terms of a Euclidean lattice.
}

\begin{document}
\maketitle

\section{Introduction}

It has long been thought \cite{Polyakov:1988md,Shaji:1990is,Fradkin:1994tt,Chen:1993cd} -- and more recently established \cite{Aharony:2011jz,Giombi:2011kc,Aharony:2015mjs,Aharony:2012nh} -- that relativistic quantum field theories in $d=2+1$ enjoy a remarkable property known as {\emph{bosonization}}.
The process of bosonization changes the statistics of particles from Bose to Fermi by flux attachment.
This is typically affected by coupling the theory to an emergent, dynamical gauge field through the introduction of a Chern-Simons term.
In addition, certain QFTs in $d=2+1$ are related by {\emph{particle-vortex duality}}, which maps bosons to bosons and fermions to fermions.
Both bosonization and particle-vortex duality map the matter content of one theory to monopole operators in another.
While the recent revival in the study of bosonization and particle-vortex duality in three dimensional systems has been spurred on by large $N$ non-Abelian gauge theories and higher spin theories, the case of Abelian dualities provides a particularly interesting framework.
From the basic prescription of Abelian bosonization via flux attachments, one can build an expanding network of dualities from a single seed relation \cite{Karch:2016sxi,Seiberg:2016gmd} (see also \cite{Murugan:2016zal}).
In this case, particle-vortex duality can be derived from bosonization.

One aspect of both the Abelian and non-Abelian cases of three-dimensional bosonization that has received little attention is the role of boundaries in the duality.
If there is any testable prediction to come from the duality, it is necessary to understand how to describe the behavior of systems with dual bulk theories in the presence of a boundary in order to make contact with quantities measurable on a finite sample.
After all, the physics of edge modes is the most easily accessible physical manifestation in quantum Hall samples.
Furthermore, including boundaries gives us one more check of the dualities which to some extent remain conjectural, even though there has been recent progress on providing a proof of the basic ``seed duality" by realizing it via a lattice construction in \cite{Chen:2017lkr}.
To understand both these points, we will study how Abelian theories on flat, half space $\Rplus$ are related by bosonization.
In particular, we will restrict our investigation to flux attachment between IR descriptions of free fermions and scalars with quartic self-coupling near or at the conformal fixed point.
We note that while we will be considering bosonization and particle-vortex duality as in \cite{Karch:2016sxi,Seiberg:2016gmd} throughout, the results obtained should be easily generalizable to include, say, non-trivial flavor symmetries \cite{Karch:2016aux}.
Despite our restriction to $\Rplus$, we believe our results generalize to curved manifolds with arbitrary boundaries so long as they are topologically trivial.

An ambitious program considering boundary conditions in 2+1 dimensional dualities has been outlined by Gaiotto in a talk over a year ago \cite{gaiotto}. In that talk, he conjectures dual pairs of boundary conditions based on constructing interfaces between a theory and its dual. Assuming that at low energies the theories decouple across the interface, an interesting web of Abelian and non-Abelian dualities emerges with subtle, non-trivial interplay of boundary conditions imposed on scalars and gauge fields. In this work, we construct a duality that agrees with one of the examples in \cite{gaiotto} and gives evidence that these conjectures -- which are based on the decoupling assumption -- are possibly true more broadly. While not attempted in the following work, it would be very interesting to flesh out the details of Gaiotto's program.

In \sn{review}, we review the status of the web of dualities in $d=2+1$ including conventions, notation, and descriptions of the theories participating in the dualities for use in all of the subsequent sections.
In \sn{half-space}, we construct the appropriate theories participating in bosonization on $\Rplus$ including possible boundary conditions and requirements for the theory to be non-anomalous.
Joining the concepts from the previous section, in \sn{dualities} we formulate the role of boundary conditions and self-consistency in describing dual theories on $\Rplus$.
Further, in \sn{lattice}, we will give evidence for the continuum duality by writing down the microscopic theory on a Euclidean three-dimensional cubic lattice.
Finally, we will conclude with an overview of the results and a discussion of future directions.
\vspace{0.5cm}

\section{Review of Abelian dualities}\label{sec:review}

To begin the analysis of Abelian dualities with boundaries, we will give a brief review of the basic players and the mechanisms that relate them \cite{Karch:2016sxi,Seiberg:2016gmd}.
Our starting point will be the two basic forms of bosonization relating a Wilson-Fisher (WF) scalar, a free Dirac fermion, and level-$k$ $U(1)$ Chern-Simons theories ($U(1)_k$ CS) living on $\Rfull$. Specifically, we begin with the ``seed'' dualities
\begin{align}\label{eq:bos1}
\text{WF scalar}~+~\text{$U(1)_1$ CS}\qquad & \longleftrightarrow\qquad  \text{Free Fermion},\\\label{eq:bos2}
\text{WF scalar}\qquad & \longleftrightarrow \qquad \text{Free Fermion}~+~\text{$U(1)_{-1}$ CS}.
\end{align}
These schematic relations are understood at the level of equating the partition functions as a function of background fields of the theories across the arrows.

In the following, we use uppercase letters to denote background gauge fields, and lowercase for a dynamical gauge fields, $\vphi$ for scalar fields, $\psi$ for Dirac fermions, and $\lambda$ for heavy Pauli-Villars regulator fields.
In what will be a necessary distinction for later application, we will denote dynamical (background) spin$_c$ valued connections with $a$ ($A$), while ordinary U$(1)$ connections will be denoted with $b ~(B),\, c~ (C)$, and so on.
The background and dynamical gauge fields are coupled through a BF-term that is defined below.
With these conventions, eqs. (\ref{eq:bos1}) and (\ref{eq:bos2}) are more precisely written respectively as
\begin{align}
		\begin{split}\label{eq:bos1-2}
		\CZ_{\text{WF+flux}}[A]&\equiv\int \CD\vphi\, \CD b ~e^{iS_{\rm WF}[\vphi,\,b] +iS_{CS}[b]+iS_{BF}[b,A]}\\ & \hspace{5.5cm}\leftrightarrow \int \CD\psi \CD\lambda  ~e^{iS_f[\psi,\lambda,\,A]} \equiv \CZ_{\text{f}}[A],
		\end{split}
\end{align}
and
\begin{align}
		\begin{split}\label{eq:bos2-2}
		\CZ_{\text{WF}}[B]\equiv\int \CD\vphi ~e^{iS_{\rm WF}[\vphi,\,B]}&\\
		&\hspace{-2.5cm} \leftrightarrow \int \CD\psi \CD\lambda \,\CD a ~e^{iS_f[\psi,\lambda,\,a] -iS_{BF}[a,B]-iS_{CS}[B]}\equiv \CZ_{\text{f+flux}}[B].
		\end{split}
\end{align}
The actions for the various matter fields participating in the above dualities are given by
\begin{subequations}
	\begin{align}\label{eq:wf}
	S_{\rm WF}[\vphi,\,B]&= \int d^3 x ~|(\partial_\mu - iB_\mu)\vphi|^2 -\alpha |\vphi|^4,\\\label{eq:dirac}
	S_{f}[\psi,\lambda,\,A] &= \lim_{m_{\lambda \rightarrow -\infty}} \, \, \int d^3x~ i\bpsi \gamma^\mu(\partial_\mu - i A_\mu)\psi  +i\bar{\lambda}\gamma^{\mu}(\partial_\mu - i A_\mu)\lambda - m_{\lambda} \bar{\lambda}\lambda\, .
	\end{align}
\end{subequations}
It is well known that a single Dirac fermion in $d=3$ has a parity anomaly, which necessitates the inclusion of the Pauli-Villars regulator in \eq{dirac} to yield a well defined fermion determinant.
Even though we are ultimately interested in the case where the regulator mass is parametrically heavy ($|m_\lambda| \rar \infty$), its effect on the theory by shifting topological terms must always be tracked -- even when $\lambda$ is integrated out.
In the literature it is common to forego writing down the regulator and instead add a $k=-\frac{1}{2}$ Chern-Simons term to the action to account for the effects of $\lambda$.\footnote{More precisely, we should note that this topological effect is the $\eta$-invariant coming from the Atiyah-Patodi-Singer index theorem \cite{Atiyah1973, Muller1994}. The precise definition will be discussed more thoroughly when the distinction is important in Sec. \sn{PV-dual}.}
We prefer to explicitly keep the regulator field around as it makes the accounting of edge modes clearer.

The actions for the level-$k$ Chern-Simons and $BF$-terms are
\begin{subequations}
	\begin{align}\label{eq:cs}
	k S_{CS}[A] &= \frac{k}{4\pi} \int d^3 x ~ \eps^{\mu\nu\rho} A_\mu\partial_\nu A_\rho,\\\label{eq:bf}
	k S_{BF}[b,A]& = \frac{k}{2\pi} \int d^3 x ~ \eps^{\mu\nu\rho} b_\mu\partial_\nu A_\rho.
	\end{align}
\end{subequations}
The normalizations in eqs.~(\ref{eq:cs}) and (\ref{eq:bf}) are chosen such that the theories with arbitrary $k\in\mathbb{Z}$ are gauge invariant in the absence of a boundary.
Taking inspiration from the microscopic description of bosonization \cite{Chen:2017lkr}, the coupling of the dynamical field $b$ to the background field $A$ can alternatively be written
\begin{align}\label{eq:no-bf}
S_{CS}[b] + S_{BF}[b,\,A]  = S_{CS}[b + A] - S_{CS}[A].
\end{align}
We will see in later sections that rewriting eqs. (\ref{eq:bos1-2}) and (\ref{eq:bos2-2}) with only Chern-Simons terms will be useful in understanding edge modes.

A few remarks are warranted before proceeding.
The statements of eqs.~(\ref{eq:bos1-2}) and (\ref{eq:bos2-2}) should to be understood at the IR fixed point.
Thus, the absence of a Maxwell term for $a$, i.e. $\frac{1}{4e^2}(da)^2$, can easily be seen because the IR limit requires $e^2\rar\infty$.
Moreover, the action for the Wilson-Fisher scalar is obtained by tuning the scalar mass $m_\vphi^2\rar 0$ and quartic coupling $\alpha\rar\infty$.
Alternatively, one can think of the Wilson-Fisher scalar by introducing an auxiliary scalar (Hubbard-Stratonovich) field, $\sigma$, such that
\begin{align}\label{eq:wf-2}
S_{WF}[\vphi,\,\sigma,\,B]=\int d^3x~\Big( |(\partial_\mu - iB_\mu)\vphi|^2-\sigma|\vphi|^2 +\frac{\sigma^2}{2\alpha}\Big).
\end{align}
Integrating out $\sigma$ produces \eq{wf}.
Treating $\sigma$ as a background field, it functions as a mass-term source.
Relating the operator insertion sourced by $\sigma$ through either of the dualities yields the map: $\sigma \leftrightarrow -\bpsi\psi$.
The way that we will interpret this map for mass deformed theories is that the scalar and fermion mass terms are mapped into one another under the duality as
\begin{equation}\label{eq:mass-dual}
\pm m_\vphi^2|\vphi|^2\qquad\leftrightarrow\qquad \mp m_\psi\bpsi\psi.
\end{equation}
Consistency of the dualities (\ref{eq:bos1-2}) and (\ref{eq:bos2-2}) for positive and negative mass deformations will be a guiding principle in what follows.

Another useful map between dualities will be between global symmetry currents.
Since we identify the global $U(1)$ symmetries on either side of the duality, it is natural to also identify the conserved currents associated with said symmetries.
For example, the duality \eq{bos1-2} implies the identification of
\begin{align}\label{eq:currents}
j^{\mu}_{\text{WF+flux}}(x)\equiv\frac{\delta S_{\text{WF+flux}}[A]}{\delta A_{\mu}(x)}\qquad\leftrightarrow\qquad j^{\mu}_{f}(x)\equiv\frac{\delta S_{f} [A]}{\delta A_{\mu}(x)}.
\end{align}
For the side of the duality with a dynamical $U(1)$ gauge field, the global $U(1)$ is associated with a flux current.
Meanwhile, the side with just matter has a global $U(1)$ that is associated with particle number.

\subsection*{Spin Considerations} \label{sec:spin}

A large portion of the subtleties involved in extending these dualities to include manifolds with boundaries comes from the differences between spin and spin$_c$ valued $U(1)$ connections.
We will now take a brief detour to review some of these concepts. The discussion here will be largely heuristic, while more mathematically oriented treatments can be found in \cite{Seiberg:2016rsg, Seiberg:2016gmd, Metlitski:2015yqa}.

Consider an arbitrary manifold, $\CM$, and turn on a background gauge field, i.e. a $U(1)$ connection $\CA$.
Suppose that we want to ask questions about the dynamics of a system of fermions on $\CM$ that couple to $\CA$.
We first must ensure that it is sensible to define the Dirac operator on $\CM$.
This requires us to define an appropriate connection, $\omega_\mu^{\,\,\,ab}$, that consistently parallel transports a local Lorentz frame over all of $\CM$, allowing us to meaningfully talk about placing a spinor anywhere on $\CM$.
An $\CM$ that admits a global definition of $\omega_\mu^{\,\,\,ab}$ is called a spin manifold.
On a spin manifold the full covariant Dirac operator is given by
\begin{align}
D_\mu  = \partial_\mu +\frac{1}{4}\omega^{\,\,\,ab}_{\mu}\gamma_a\gamma_b + i \CA_\mu.
\end{align}

However, certain topological constraints imply that not every manifold admits a global definition of $\omega_{\mu}^{\,\,\,ab}$.
The topological obstruction to defining $\omega$ everywhere on $\CM$ can be compensated by a non-standard choice for quantization of $\CA$:
\begin{equation}\label{eq:spin-charge}
\frac{1}{2\pi} \int_{\Sigma} d\CA\in 2\mathbb{Z},
\end{equation}
where $\Sigma$ is an oriented co-dimension 2 surface in $\mathcal{M}$.
Within this quantization scheme, the covariant Dirac operator
\begin{align}
D_{\mu}^{(n)}=\partial_\mu + \frac{1}{4}\omega_\mu^{\,\,\,ab}\gamma_a\gamma_b+in\CA_{\mu}
\end{align}
is well defined for odd $n$.
An $\CM$ whose topological obstruction to a global definition of the Dirac operator is compensated by the unusual quantization of $\CA$ is called a spin$_c$ manifold, and the $\CA$ obeying \eq{spin-charge} will be referred to as a spin$_c$ valued connection.

Further, we can impose \eq{spin-charge} even if the manifold admits a global definition of $\omega_{\mu}^{\,\,\,ab}$, which implies that spin and spin$_c$ valued connections can be defined spin manifolds .
Thus since $\Rfull$ and $\Rplus$ are spin manifolds, the distinction that we must make is at the level of fermions being coupled to either a spin or spin$_c$ valued connection.

The restriction to odd $n$ gives rise to the spin-charge relation of condensed matter physics; particles with integer spin have even charge and half-integer spin have odd charge.
While this does not appear to be a fundamental law of nature, it is believed to be valid for systems made up of protons, electrons and other charged (quasi-)particles.
This motivates the distinction between spin and spin$_c$ valued connections in our notation and further implies that our background field appearing in \eq{bos1-2} is spin$_c$ \cite{Seiberg:2016rsg}.

As an example of how this distinction can enter into the seed dualities, consider pure a $U(1)_1$ theory with spin valued connection, $b$, on $\CM=T^3$.\footnote{This discussion follows Appendix B of \cite{Seiberg:2016gmd} where more details can be found.}
Further, consider that $\CM$ is the boundary of a four-dimensional manifold $X$.
Upon quantization, we find that there is just one state such that the path integral is
\begin{equation}
\mathcal{Z}=\int \mathcal{D} b ~ e^{iS_{CS}[b]} = e^{-i\Omega}
\end{equation}
where $\Omega$  denotes ``framing anomaly''
\begin{equation}\label{eq:grav_cs}
\Omega=2\int_{\partial X}\text{CS}_{\text{g}}=\frac{1}{96\pi}  \int_{X}\mathcal{R}\wedge \mathcal{R},
\end{equation}
and $\text{CS}_{\text{g}}$ is the gravitational Chern-Simons term.

If the same $U(1)$ theory with dynamical gauge field $b$ is defined with respect to a background spin$_c$ structure with connection $A$ on $\CM$, we must couple our dynamical $U(1)$ field $b$ to the background connection $A$ through a $BF$ term.
As with the previous case, there is only one state, and the theory is uniquely determined.
The difference is that the partition function evaluates to
\begin{equation}
\mathcal{Z}[A]=\int \mathcal{D} b ~ e^{iS_{CS}[b]+iS_{BF}[b,A]} = e^{-i\Omega-i\frac{1}{4\pi}\int AdA}.
\end{equation}
Accounting for these extra terms will prove to be a useful guiding principle for keeping track of edge modes across the duality.

\section{Theories on half-space}\label{sec:half-space}

Now that we have reviewed the basics of the standard Abelian dualities, we are in a position to address the subtleties associated with the theories on the half-space, $\Rplus$.
We will explore the space of boundary conditions consistent with eqs. (\ref{eq:dirac}) or (\ref{eq:wf}) defined on $\Rplus$.

To do so, we must remind ourselves of how to be honest about boundary conditions in field theories.
Consider a theory with action $S$ defined on the manifold $\mathcal{M}$ with boundary $\partial\mathcal{M}$.
By taking the variation $\delta S$ we will find two classes of terms
\begin{equation}
\delta S=\delta S_{\text{bulk}}+\delta S_{\text{bdry}}=\int_{\mathcal{M}}\delta\mathcal{L}_{\text{bulk}}+\int_{\partial\mathcal{M}}\delta\mathcal{L}_{\text{bdry}}.
\end{equation}
The bulk part of the action is still extremized by the classical equations of motion, and consistency of the variation amounts to choosing conditions on the field configurations such that $\delta S_{\text{bdry}}$ vanishes as well.
In the classical limit of the theory, the field configuration that satisfies the equations of motion should also satisfy boundary conditions.
In the full quantum theory this is not necessarily the case. One way to proceed is by manually restricting the space of allowed field configurations by inserting delta funtions in the path integral which impose the desired boundary conditions.
%
%
This method excludes fluctuations where $\delta S_{\text{bdry}}\ne0$. Alternatively, we could do the path integral over all boundary field configurations. In that case the boundary conditions would only be obeyed by the dominant field configurations in the path integral--those which extremize the action.
%
%
Below we will see that for all fields we consider there will be multiple boundary conditions which satisfy $\delta S_{\text{bdry}}=0$.
The boundary conditions will be chosen such that the theory remains non-anomalous and we keep the global symmetries on either side of the duality consistent.

In addition to the field conventions listed above, we will take coordinates on $\Rplus$ to be $\{t,\,x,\,y\}$ where $t,\,x\in (-\infty,\,\infty)$ and $y\geq 0$.
The boundary of $\Rplus$ is the surface at $y=0$.
Indices $i,\,j$ will be used to denote coordinates on the boundary and $\mu,\,\nu$ in the bulk.

\subsection{Boundary conditions}\label{sec:neumann}

Applying the above approach to \eq{wf}, we take the theory defined on $\Rplus$ by fiat and vary such that
\begin{align}
\delta S_{WF}[\vphi, B] = \ldots + \int_{y=0} d^2x\,\Big(\delta\vphi^\dagger D_y\vphi +\delta\vphi D_y\vphi^\dagger\Big)
\end{align}
where ``$\ldots$'' contains bulk terms which vanish on-shell.
This implies that both Dirichlet $\delta\vphi|_{y=0}=0$ and Neumann $D_y\vphi|_{y=0} =0$ are valid boundary conditions.

Now consider the boundary conditions for a Dirac fermion.
We write the Dirac fermion \eq{dirac} evaluated on $\Rplus$ in terms of left and right handed components:
\begin{align}
\psi = \begin{pmatrix}\psi_+\\ \psi_-\end{pmatrix}, \qquad \text{i.e.}\quad\psi_\pm&=P_\pm\psi \qquad \text{with} \qquad P_{\pm} = \frac{\mathds{1}\pm\gamma^y}{2},
\end{align}
and $\gamma^y$ is the gamma matrix in the direction perpendicular to the boundary.
$\gamma^y =i\gamma^t\gamma^x$ is the `$\gamma_5$' in the boundary theory.
Now, on $\Rplus$  the terms in \eq{dirac} that depend on $\psi$ in this language read
\begin{align}\label{eq:chiral-bdry}
	\begin{split}
	S_{f}[\psi,\lambda,\,A]&= \int_{\Rplus}d^3x~ \Big(i\bpsi_+\Dslash_{A_i}\psi_+ +i\bpsi_-\Dslash_{A_i}\psi_- +\bpsi_-
	A_y\psi_+-\bpsi_+A_y\psi_-\\
	&\hspace{2.5cm}+\frac{i}{2}(\bpsi_-\partial_y\psi_+-\partial_y\bpsi_-\psi_+ +\partial_y\bpsi_+\psi_- -\bpsi_+\partial_y\psi_-)\Big) + \ldots
	\end{split}
\end{align}
where the ellipses denote the terms that only depend on the Pauli-Villars regulator field and $\Dslash_{A_i} \equiv \gamma^i(\partial_i - i A_i)$.

The boundary terms generated by the variation of \eq{chiral-bdry} are
\begin{align}\label{eq:fermion-bcs}
\delta S_{f}[\psi,\,A] =\ldots+ \int_{y=0} d^2x~\big(\psi_-\delta\bpsi_+ - \psi_+\delta\bpsi_- +\bpsi_- \delta \psi_+ -\bpsi_+\delta\psi_-\big).
\end{align}
We can consistently impose Dirichlet boundary conditions on {\emph{either}} of the chiral components,\footnote{We will always take $P_{\pm}\psi|_{y=0}$ to imply the corresponding relation on the conjugate field, namely $\bar{\psi}P_{\mp}|_{y=0}=0$.}
\begin{align}\label{eq:ferm-bc}
\psi_+|_{y=0} = 0\qquad\text{or}\qquad\psi_-|_{y=0} = 0.
\end{align}
However, choosing {\emph{both}} $\psi_+|_{y=0} = 0$ and $\psi_-|_{y=0} = 0$ over-constrains the equations of motion at the boundary \cite{Csaki:2003sh}.
Either choice in \eq{ferm-bc} leaves behind a chiral edge mode as seen in the current running parallel to the boundary, $j^i_\psi = \bpsi\gamma^i\psi$.
In \sn{bosonization}, we will explore how requiring a non-anomalous theory forces us to choose one boundary condition over the other.

Since the action for the Pauli-Villars regulator fields is identical to that for the Dirac fermions, the analysis above applies in kind.
In particular, we apply chiral boundary conditions on the Pauli-Villars regulator as well, i.e.
\begin{align}\label{eq:pv-bc}
\lambda_+|_{y=0} = 0\qquad\text{or}\qquad\lambda_-|_{y=0} = 0.
\end{align}
In what follows, we will show boundary condition of the Dirac fermion and Pauli-Villars field are related.
In order to keep track of which boundary condition we are imposing on the two fields, we introduce a superscript
$S_f^{\pm}[\psi,\lambda, \, A]$ to indicate imposing the boundary conditions $\left . \psi_{\mp} \right |_{y=0}=\left . \lambda_{\pm} \right |_{y=0}=0$.

We will use similar notation for the time reversed fermion actions $\bar{S}_f^{\mp}[\psi,\lambda,A]$ with the superscript indicating the type of boundary conditions imposed.
Note that the time reversed version of $S_f^+$ is $\bar{S}_f^-$ and vice versa.
In addition, we should keep in mind that $S_f^{\pm}$ itself was defined with a large negative Pauli-Villars mass, and since fermion mass terms are time-reversal odd, $\bar{S}_f^{\pm}$ is defined with a large positive Pauli-Villars mass.
This means that $\bar{S}_f$ can be thought of as coming with a $k=+\frac{1}{2}$ Chern-Simons term rather than $k=-\frac{1}{2}$.

Next, consider the possible boundary conditions for our dynamical gauge fields.
To constrain such fields, we will consider the action at the level of the microscopic description in which the Maxwell term is still dominant.
Upon variation, we find
\begin{align}\label{eq:maxwell-bcs}
\delta S_{\text{Maxwell}}[b] =\ldots-\frac{1}{e^{2}} \int_{y=0} d^2x~F^{yi}\delta b_{i}.
\end{align}
with $F_{yi}=\partial_{y}b_{i}-\partial_{i}b_{y}$.
Once more, we see we can impose either Dirichlet or Neumann boundary conditions.
The former requires the variation along the boundary to vanish, i.e. $b_i=0$.
Neumann boundary conditions require the field strength adjacent and oriented perpendicular to the boundary be flat, $F_{iy}=0$.

Lastly, we will consider the boundary conditions for a level-$k$ Chern-Simons term.
Such terms will only come up in the IR limit of the dualities.
Varying \eq{cs} gives
\begin{align}
k ~ \delta S_{\text{CS}}[b] =\ldots +\frac{k}{4\pi} \int_{y=0}d^2x~\eps^{ij}b_i\delta b_j.
\end{align}
While we could impose $b_t = 0$ or $b_x = 0$ at the boundary, requiring the general, sufficient condition that
\begin{align}\label{eq:flux-bc}
(b_t-v_b b_x)|_{y=0} =0,
\end{align}
makes the boundary physics clear in the context of eqs. (\ref{eq:bos1-2}) and (\ref{eq:bos2-2}).
That is, we maintain a chiral edge mode with velocity $v_b$ and chirality set by sgn$(v_b)$.
In order for the boundary kinetic term to be positive definite, the velocity must be chosen such that $v_b k>0$ \cite{Fradkin:1991nr}.
In what follows we will be mostly interested in relativistic theories fixing the magnitude $|v_b|=1$.
Since a gauge transformation of \eq{cs} also produces a boundary term, any gauge choice that we make must be consistent with \eq{flux-bc}.
The simplest solution is to promote the boundary condition to a gauge fixing condition, i.e. we let $(b_t-v_b b_x)=0$ in the bulk as well.
As we will see in the next section, the freedom to choose $v_b$ is actually tied to the choice of fermionic boundary conditions.
The consistency requirement on the sign of $v_b$ will then pick a preferred fermionic boundary condition, which we will hardwire into the path integral.


In this section, we have seen that there are multiple choices of boundary conditions for all of the fields in our theories.
However, the choices will be constrained by requiring the theory to be non-anomalous and that the global symmetries on either side of the duality match.

\subsection{Boundary modes and anomalies}\label{sec:DWF}

The discussion of the previous subsection will prove sufficient to study the duality between the conformal field theories related by bosonization.
However, to check the consistency of our dualities under deformations, we will also be interested in adding mass gaps to the theories on $\Rplus$.
Before formulating dualities like eqs.~(\ref{eq:bos1-2}) and (\ref{eq:bos2-2}) with boundary conditions, we will highlight additional subtleties in gapped phases in the presence of a boundary.

Our main concern in this section is the possible existence of domain wall fermions (DWFs) and their interplay with anomalies.\footnote{Strictly speaking, our ``domain wall'' is really the boundary of our material, but we will continue to use this slight abuse of vocabulary.}
DWFs are typically discussed in the context of Dirac fermions defined on $\Rfull$ with a spatially varying mass term -- specifically, a mass term that changes sign across an interface.
 But the same basic construction also allows us to look for the existence of massless boundary modes on $\Rplus$.
A massless chiral mode localized on the boundary will exist when the mass profile leaves
\begin{equation}
\label{eq:DWF-cond}
\xi\left(y\right)=e^{\pm\int_{0}^{y}dy^\prime\,m\left(y^\prime\right)}
\end{equation}
finite for all $y\in\Rplus$ \cite{Kaplan:1992bt}.
Unlike the DWF descending from the construction on $\Rfull$, any constant, non-zero mass profile ($m(y)=m$) in \eq{DWF-cond} yields a normalizable zero mode for a fermionic theory on $\Rplus$.
The chirality of the DWF is set by the sign of the mass: sgn$(m)=+1$ gives a left-mover and  sgn$(m)=-1$ a right-mover.
In either case, the chiral current is not conserved, and so the boundary theory on its own is anomalous.
While this is not necessarily an inconsistency in the case when the fermion number is not gauged, we are only interested in theories in which our global symmetry currents are in fact conserved and so can be consistently coupled to background fields.\footnote{Dualities between theories which have non-vanishing boundary anomalies for global symmetries can also be formulated, as long as the anomalies on both sides of the duality agree. We do not consider such dualities in this work, but they have been outlined in \cite{gaiotto} along with the theories we consider.}

It has long been known that a level-$k$ Chern-Simons in the bulk can precisely account for the anomalous chiral modes living on the defect so long as they satisfy the relation
\begin{align}
\label{eq:CH-mech}
k = n_+ - n_-,
\end{align}
where ($n_+$) $n_-$ are the number of (right-) left-moving modes.
More precisely, the nonzero anomaly of the bulk Chern-Simons term under gauge transformations of its associated gauge field can be exactly compensated by the axial anomaly of chiral edge movers on the boundary.
This is know as the Callan-Harvey mechanism \cite{Callan:1984sa}.

In addition to the chiral anomaly, there is also a framing anomaly of such edge theories which arises under diffeomorphism transformations.
There is a condition analogous to the Callan-Harvey mechanism which accounts for anomalies associated with diffeomorphism transformations of the gravitational Cherm-Simons terms we will consider.
In particular, a manifold $\mathcal{M}$ with a boundary will not be diffeomorphism invariant unless the theory satisfies
\begin{equation}
\label{eq:gCH-mech}
k_{\Omega} = \frac{1}{2} \left(n^{\text{MW}}_+-n^{\text{MW}}_-\right)
\end{equation}
where $k_{\Omega}$ is the coefficient of gravitational Chern-Simons term, $i\Omega$ of \eq{grav_cs}, and $n^{\text{MW}}_{\pm}$ the number of right- and left-moving Majorana-Weyl fermions, respectively.
Fortuantely, a single chiral Dirac fermion is equivalent to two Majorana-Weyl fermions, i.e. $n_{\pm}=2n^{\text{MW}}_{\pm}$ \cite{Seiberg:2016gmd}.
Hence, so long as $k=k_{\Omega}=\pm 1$, a single chiral fermion can render the theory non-anomalous for both the chiral and framing anomalies.
In what follows, our calculations will be organized such that keeping track of \eq{CH-mech} is completely equivalent to \eq{gCH-mech}.

We will see that requiring our theories to be non-anomalous -- such that \eq{CH-mech} is satisfied -- arranges for us the pieces laid out above into a working conjecture for Abelian dualities with a boundary.
Furthermore, this counting will naturally appear as an organizational tool in the lattice construction in later sections.

Let us now take account of the possible edge modes that can appear in the context of bosonization dualities.
To start, we will only consider matter fields.
The scalars will never give rise to a chiral edge mode.
For gapped fermions, we naturally get DWFs subject to the boundary conditions of \eq{ferm-bc}, which select any possible surviving edge mode.
As hinted by \eq{CH-mech} these DWFs are intimately connected to Chern-Simons terms.

By the same reasoning that our gapped fermions give rise to DWFs,  so too do the Pauli-Villars fields.
We will always take the boundary conditions on the Pauli-Villars regulators to kill off the would-be DWF.
If we do not kill off the Pauli-Villars DWF, this would give us massless ghosts localized to the boundary.
This would be orthogonal to the Pauli-Villar's field original purpose, which was to regulate high energy degrees of freedom giving rise to the parity anomaly.
%

Consider a spin$_c$ valued connection, $A$, coupled to a heavy Dirac fermion, $\chi$, and a heavy Pauli-Villars regulator, $\lambda$, with positive masses.
Here we will take $A$ to be a background field, but analogous results hold for dynamical spin$_c$ valued connections up to potential boundary conditions which we will discuss later.
The effective action generated by integrating out a heavy Dirac fermion is $iS_{CS}[A] + i\Omega$.
Furthermore from \eq{DWF-cond}, $\chi$ gives rise to a DWF of positive chirality, and so we can satisfy \eq{CH-mech} by imposing $\chi_-|_{y=0} = 0$ to leave the DWF unaffected.
The same DWF is precisely the edge mode we also need to account for the framing anomaly.
To remove the DWF associated with the Pauli-Villars regulator, we impose $\lambda_+|_{y=0} = 0$.
Analogous results follow choosing negative mass Dirac fermions and Pauli-Villars regulators with a flipped Chern-Simons level and the opposite boundary conditions.
Choosing the signs of the fermion and Pauli-Villars masses to be anti-aligned, the Chern-Simons terms cancel.
Furthermore, both the fermion and Pauli-Villars boundary conditions prevent any DWFs from arising.
As promised, for $k=\pm 1$ only one of the two possible fermionic boundary conditions yields a theory consistent with \eq{CH-mech}.

Returning to the IR boundary conditions on the gauge fields, we saw the Chern-Simons term gave us a chiral edge mode whose handedness was set by the sign by the velocity in \eq{flux-bc} and hence by $k$.
If this Chern-Simons term is generated by integrating out a massive fermion, the bosonic chiral edge mode from the gauge field can be understood as a $1+1$-dimensional bosonized DWF.
Thus, the IR physics still retains some memory of the microscopic picture due to the gapless chiral edge mode furnished by the underlying DWF, which appropriately accounts for the anomalies.
Together the massive fermions, Maxwell term for the gauge field, and chiral edge mode give a complete microscopic picture of the theory.
This implies that \eq{flux-bc} emerges from the boundary conditions imposed on the microscopic fermions.

In fact, we would like to promote this to an operating principle for how to deal with Chern-Simons terms when analyzing theories in the presence of boundaries.
We want to view {\it all} spin$_c$ and gravitational Chern-Simons terms as being generated by integrating out massive fermions.
This is the easiest way to get a consistent microscopic picture accounting for all the resulting boundary modes and anomaly inflows.
In particular, this means we will have the following view of the Chern-Simons terms appearing in Abelian bosonization:
\begin{quote}
\textbf{$+iS_{CS}[A] + i\Omega$ Chern-Simons terms:} Pauli-Villars regulator and a free fermion with $m_{\lambda},\,m_{\chi}>0$ and the $\chi_-|_{y=0} = 0$ and $\lambda_+|_{y=0} = 0$ boundary conditions.

\textbf{$-iS_{CS}[A] - i\Omega$ Chern-Simons terms:} Pauli-Villars regulator and a free fermion with $m_{\lambda},\,m_{\chi}<0$ and the $\chi_+|_{y=0} = 0$ and $\lambda_-|_{y=0} = 0$ boundary conditions.
\end{quote}
The signs of the masses of the fermion and Pauli-Villars fields and their appropriate boundary conditions are completely determined by the sign of the Chern-Simons level.
We will use this microscopic description both for Chern-Simons terms for dynamical spin$_c$ fields and for Chern-Simons terms associated with background spin$_c$ valued connections.
For clarity, we will denote the fermions that appear in eqs. (\ref{eq:bos1-2}) and (\ref{eq:bos2-2}) as $\psi$ and refer to them as ``dynamical'', while  ``fiducial'' fermions $\chi$ refer to the microscopic description of the Chern-Simons term.
More explicitly, we will view every Chern-Simon terms as arising from
\begin{equation}
\label{eq:FF_expr}
e^{\pm i S_{CS}[A] \pm i \Omega} = \int \CD\chi\,\CD\lambda~ e^{i S_{ff}^{\pm} [\chi,\,\lambda,A]},
\end{equation}
where
\begin{equation}
	S^{\pm}_{ff}[\chi,\lambda,A] = \lim_{|m_{\chi}|,|m_{\lambda}| \rightarrow \infty} \, \, \int_{\Rplus}d^3x~ \Big(i\bar{\chi} \Dslash_A \chi \mp |m_{\chi}| \bar{\chi} \chi +i\bar{\lambda} \Dslash_A \lambda \mp |m_{\lambda}| \bar{\lambda}\lambda\Big).
\end{equation}
The superscript on fiducial fermion action denotes the sign of the fermion and Pauli-Villars masses\footnote{This is to be contrasted with our definition of $S_f^{\pm}[\psi,\lambda,A]$ for the dynamical fermions. The latter were massless to begin with and we always took the Pauli-Villars mass to be negative and large. The superscript in that case only referred to the boundary conditions.} as well as the corresponding boundary conditions, $\chi_\mp|_{y=0}=\lambda_\pm|_{y=0} = 0$.
As usual, we have chosen the convention that the fermionic mass term appears generically as  $V(\psi)=+m_{\psi}\bpsi\psi$.

The only difference between dynamical and background spin$_c$ valued connections is the possibility of imposing boundary conditions on the former.
Since Dirichlet boundary conditions set the gauge field at the boundary to zero, imposing them will eliminate anomalous current flow onto the boundary from a dynamical Chern-Simons term.
Hence, we do not need to put any additional chiral boundary modes to compensate for such currents.
However, employing Dirichlet boundary conditions changes the boundary gauge symmetry to a global symmetry; thus, introducing a second global $U(1)$ symmetry into the theory.
On the dual side, new boundary localized matter has to be added to account for this enhanced global symmetry.
In this work, we will only consider Neumann boundary conditions on the dynamical gauge fields so that \eq{CH-mech} needs to be satisfied for all types of gauge fields. Additional dualities with Dirichlet boundary conditions on gauge fields have been outlined in \cite{gaiotto}.

We will see that the boundary modes associated with the fiducial fermions will be crucial in developing a consistent picture of boundary modes.
This is particularly interesting when the Chern-Simons terms involved describe only background fields. In this case the fiducial fermion still can contribute massless boundary modes, even though the Chern-Simons term does not involve any fluctuating fields.
From the point of view of the low energy theory it appears that these fermionic boundary modes have to be added ``by hand" in order for the duality to hold.

\section{Dualities including boundaries}\label{sec:dualities}

We now turn to establishing three-dimensional bosonization and particle-vortex duality in the presence of a boundary.
Our starting point is the conjecture that dualities \eqref{eq:bos1} and \eqref{eq:bos2} are valid on $\Rplus$ provided the boundary conditions are correctly applied to dynamical and fiducial fermions.
From this conjecture, we will also be able to establish a web of Abelian dualities -- i.e. scalar-vortex and fermion-QED$_3$ -- in the presence of boundaries.
%
%
The derivation will give us a setting to establish checks between chiral degrees of freedom on the boundary and Chern-Simons levels such that \eq{CH-mech} is satisfied at every step of the way.
All partition functions in this and subsequent sections are understood to be defined on the half-space and distinct from their full-space equivalents.

\subsection{Bosonization}\label{sec:bosonization}

\subsubsection*{Scalar+Flux = Fermion}

Our conjecture for the form of the seed duality with a boundary starts with rewriting the flux attachment to Wilson-Fisher scalars using \eq{no-bf},
\begin{align}\label{eq:wf-flux-4}
\CZ_{\text{WF+flux}}[A] = \int \CD\vphi\,\CD b\, e^{iS_{WF}[\vphi,b] +i S_{CS}[b+A]-iS_{CS}[A]}.
\end{align}
In this form, the coupling of the statistical gauge field $b$ to the background $A$ can be understood entirely in terms of the microscopic fiducial description via heavy fermions:
\begin{align}\label{eq:wf-flux-4-2}
\CZ_{\text{WF+flux}}[A] = \int \CD\vphi \, \CD b\prod_{j=1,2}\CD\chi_j \CD\lambda_j \, e^{iS_{WF}[b] +i S_{ff}^+ [\chi_1,\,\lambda_1, b+A] +i S_{ff}^-[\chi_2,\,\lambda_2,\,A]},
\end{align}
where once again the superscripts are chosen such that they generate the corresponding Chern-Simons terms appearing in \eq{wf-flux-4}.
Implicit in the above expression is the fact the gravitational Chern-Simons terms coming from each of the fiducial fermions cancel,
\begin{equation}
\left(iS_{CS}[b+A]+i\Omega\right)+\left(-iS_{CS}[A]-i\Omega\right)=iS_{CS}[b+A]-iS_{CS}[A].
\end{equation}
This particular combination of Chern-Simons terms will be used many times in what follows.

We should reemphasize that this rewriting has actual content in the case of a theory with boundary: Even though $A$ is a non-dynamical background gauge field, $S_{ff}^-[\chi_2,\,\lambda_2,\,A]$ will give rise to massless chiral boundary modes associated with the fiducial fermion $\chi_2$ despite working in the $|m_{\chi_2}|\rar\infty$ limit.
As noted above, from the perspective of the coarse-grained, Chern-Simons formulation of the theory in \eq{wf-flux-4} these gapless edge modes appear to be added by hand.

\begin{table}
\protect\begin{centering}
\begin{tabular}{ccc}
\hline 
 & \textbf{Scalar $+$ flux} & \textbf{Fermion}\tabularnewline
\hline 
\textbf{Boundary conditions} & $\protect\vphi=0$ & $\chi_{+}=0$\tabularnewline
 & $\protect\partial_{y}b_{i}-\protect\partial_{i}b_{y}=0$ & \tabularnewline
\hline 
\textbf{Additional edge modes} & Left-mover coupled to $A$ & None\tabularnewline
 & Right-mover coupled to $b+A$ & \tabularnewline
\hline
\end{tabular}\protect
\par\end{centering}
\protect\caption{Summary of boundary conditions and additional edge movers for \eq{bos1-4}.\label{tab:seed1_tab}}

\end{table}

The fermionic side of the duality \eq{bos1-2} does not need any additional work:
It is already in a form that makes the chiral edge modes obvious.
We can simply apply the chiral boundary conditions on dynamical fermions ($\psi_{+}|_{y=0}=0$) and Pauli-Villars regulator ($\lambda_{-}|_{y=0}=0$).
Our conjecture is then that
\begin{align}\label{eq:bos1-4}
\begin{split}
\CZ_{\text{WF+flux}}[A]\equiv & \int \CD\vphi ~\CD b~\prod_{j=1,2} \CD\chi_j\,\CD\lambda_j~ e^{iS_{WF}[b] +i S^+_{ff} [\chi_1,\,\lambda_1, b+A] +i S^-_{ff}[\chi_2,\,\lambda_2,\,A]} \\ & \hspace{5cm}\leftrightarrow \int \CD\psi~\CD\lambda \,e^{iS^-_{f}[\psi,\lambda,\,A]}\equiv\CZ_{\text{f}}[A]
\end{split}
\end{align}
holds as an equivalence at the conformal point.
Additionally, we choose the dynamical gauge field to obey Neumann boundary conditions, $(\partial_y b_i - \partial_i b_y)|_{y=0} = 0$, and the scalar to obey the Dirichlet condition, $\vphi|_{y=0} = 0$.
These results are summarized in Table \ref{tab:seed1_tab}.

In order to establish some guiding principle for the conjectured duality of CFTs, we can gap both theories and track whether our putative equivalence holds for positive and negative mass deformations.
We will see the boundary conditions in our conjecture naturally arise by requiring the theory to be non-anomalous and have consistent global symmetries.
With the correspondence of signs between fermion and scalar mass terms in the original bosonization duality in \eq{mass-dual} and the convention we've already chosen for fermions, the potential for the scalars is $V(\vphi) = -m_{\vphi}^2|\vphi|^2+\alpha|\vphi|^4$.
We should find consistent dualities between theories in the bulk and on the boundary for positive and negative mass deformations away from the CFT.

Let us start with the free fermion side of \eq{bos1-2}.
Making the mass deformation explicit, the action is given by the replacement
\begin{align}
 S_{f}^+[\psi,\lambda,A] \rightarrow S_{f}^+[\psi,\lambda,A] -m_{\psi}\bpsi\psi
\end{align}
where $\psi_{+}|_{y=0}=0$.
In the IR limit of the theory, integrating out the massive degrees of freedom of the fermion yields
\begin{equation}
\label{CS-level}
S_{f} = - \frac{1}{2}\left(1-\text{sgn}(m_\psi)\right) (iS_{CS}+i\Omega)
\qquad \text{(IR Limit)}.
\end{equation}
When the Pauli-Villars field and the fermion have the same sign of mass, corresponding to a $-iS_{CS}[A]-i\Omega$ Chern-Simons term, we need a single left-moving chiral edge mode to account for the anomalous term in order for this to be consistent with \eq{CH-mech}.
Since $m_{\psi}<0$, the DWF which arises from our analysis of \sn{DWF} is exactly the anomaly cancelling edge mode we need.
If instead we had imposed the condition $\left.\psi_-\right|_{y=0}=0$, then this would have suppressed the DWF.
Hence, if we demand a non-anomalous theory, we are forced into choosing $\left.\psi_+\right|_{y=0} = 0$.

We should now check to make sure everything is consistent for $m_{\psi}>0$.
In this case we get no ordinary or gravitational Chern-Simons terms and $\psi$'s mass profile naturally gives rise to a right-moving DWF.
It seems like we are in trouble.
Fortunately, applying $\left.\psi_+\right|_{y=0}=0$ prevents any right-movers on the boundary.
We are thus left with no chiral edge modes and \eq{CH-mech} is satisfied for both signs of $m_{\psi}$.

For the Wilson-Fisher scalar with flux, introducing a mass deformation $m_\vphi^2<0$ with our conventions for $V(\vphi)$ gives an overall positive mass term that corresponds to a gapped scalar.
Flowing to the IR, the only term with $b$ dependence is $iS_{CS}[b+A]$.
As reviewed in above and in appendix B of \cite{Seiberg:2016gmd}, this theory is completely determined by its framing anomaly and thus equal to $-i\Omega$.
This results in an overall $-iS_{CS}[A]-i\Omega$ Chern-Simons term, consistent with the fermionic side when $m_{\psi}<0$.

We should also check that the anomaly inflow condition \eq{CH-mech} is still satisfied on this side of the duality.
It is here where our microscopic description of the Chern-Simons term in \eq{wf-flux-4-2} will be important.
Integrating out $b$ caused the first Chern-Simons term to vanish leaving behind $-i S_{CS}[A] - i \Omega$.
From the micrscopic perspective, this can be viewed as the condition
\begin{equation}\label{eq:triv_1}
\int \CD\chi~\CD\lambda~\CD b ~ e^{iS_{ff}^{\pm}[\chi,\lambda,b+A]}=1.
\end{equation}
That is, the fiducial fermions provide no ordinary or gravitational Chern-Simons terms as well as no corresponding edge movers.
Per our prescription, the remaining fiducial fermion associated with $-i S_{CS}[A] - i \Omega$ has the correct mass profile and boundary condition such that it contains a left-moving DWF.
Thus, \eq{CH-mech} is satisfied.

To complete our discussion of massive phases we need to check that everything is consistent when $m_{\vphi}^2>0$.
This gives a negative mass squared term in $V(\vphi)$, spontaneously breaking the emergent $U(1)$ in the scalar theory.
This kills off the Chern-Simons term for $b$, and so integrating out $\vphi$ and $b$ leave behind no Chern-Simons terms.
As expected, this means that the IR theory in the Higgs phase is identical to the `vacuum' region.
When $b=0$, the edge modes of the the fiducial fermions associated with $iS_{CS}[b+A]$ and $-iS_{CS}[A]$ have the same gauge coupling but opposite chiralities, and hence cancel one another.
Since no Chern-Simons terms or fermions are left behind, there are no possible chiral modes that can arise and make this theory anomalous.
Hence, we have found a consistent story for the duality on either side of the mass deformation.

That last step is to see if the scalar boundary conditions is constrained.
To do so, we rely on our identification of global symmetry currents on either side of the duality, \eq{currents}.
For this purpose, it becomes useful to reinterpret the cancellation of the anomaly from \eq{CH-mech} in a slightly different, but equivalent, language.
The Chern-Simons term of the bulk is anomalous on its own under the global $U(1)$ topological symmetry because the corresponding current has a nonzero divergence at the boundary.
This seems to imply that the symmetry is broken at the boundary.
However, the Chern-Simons anomaly is compensated via the axial $U(1)$ symmetry of the DWFs, and hence the theory is non-anomalous under a simultaneous topological $U(1)$ transformation in the bulk and the axial $U(1)$ transformation on the DWFs.
If the two symmetries are identified, the global topological $U(1)$ symmetry is restored on the boundary by the transformation of the DWFs and is unbroken everywhere.
This is in agreement with the fermion side of the duality where the global $U(1)$ symmetry of particle number is unbroken in the bulk and on the boundary.

Returning to the constraints on the boundary condition of the scalar, recall that the equations of motion for the scalar and Chern-Simons term tie the matter current to the topological current,
\begin{align}\label{eq:flux-scal}
j^{\mu}_{\text{flux}}\equiv \frac{k}{2\pi}\epsilon^{\mu\nu\rho}\partial_{\nu}b_{\rho}=-j_{\text{scalar}}^{\mu}.
\end{align}
Here, $j_{\text{scalar}}^{\mu}$ is the usual scalar matter current and we have temporarily set the background fields to zero.
However, as we have argued above, on the boundary it is not the flux which accounts for the topological $U(1)$ symmetry, but the DWFs.
Hence, we should have $j^{i}_{\text{flux}}|_{y=0}=0$ and by \eq{flux-scal} should also take $j_{\text{scalar}}^{i}|_{y=0}=0$.
Such a condition on the scalar current can only be achieved by Dirichlet boundary conditions, $\vphi|_{y=0}=0$.
Dirichlet boundary conditions are usually referred to as the ``ordinary transition'' boundary conditions of the $O(2)$ Wilson-Fisher fixed point. See \cite{Liendo:2012hy} for a recent discussion.

The above constructions leads us to conjecture what happens to the DWFs at the conformal fixed point:
As the mass deformation becomes smaller, according to \eq{DWF-cond} the DWF becomes less and less localized to the boundary.
In the massless limit, the DWF recombines with a DWF of opposite chirality living on -- in the case of a finite interval $y\in [0,\,L]$ -- the other boundary.
Note that on the semi-infinite interval that we have used for $\Rplus$, the oppositely chiral fermion is not explicitly seen as the boundary condition at $y=L$ is replaced by a condition on the asymptotic behavior of the matter fields.
At the conformal fixed point, we then have an ordinary Dirac fermion which lives in the bulk.

\subsubsection*{Fermion+Flux = Scalar}

\begin{table}
\protect\begin{centering}
\begin{tabular}{ccc}
\hline 
 & \textbf{Fermion $+$ flux} & \textbf{Scalar}\tabularnewline
\hline 
\textbf{Boundary conditions} & $\chi_{-}=0$ & $\protect\vphi=0$\tabularnewline
 & $\protect\partial_{y}a_{i}-\protect\partial_{i}a_{y}=0$ & \tabularnewline
\hline 
\textbf{Additional edge modes} & Left-mover coupled to $a+B$ & None\tabularnewline
 & Right-mover coupled to $a$ & \tabularnewline
\hline 
\end{tabular}\protect
\par\end{centering}
\protect\caption{Summary of boundary conditions and additional edge movers for \eq{bos2-4}.\label{tab:seed2_tab}}
\end{table}

Having established a set of conventions in the first seed duality in the presence of a boundary, we can carry the above notation through into the second seed duality.
Our conjecture is that
\begin{align}\label{eq:bos2-4}
\begin{split}
\CZ_{\text{f+flux}}[B]\equiv&\int \CD\psi~\CD a~\CD\lambda_0~\prod_{j=1,2} \CD\chi_j\,\CD\lambda_j ~e^{iS_{f}^+[\psi,\,\lambda_0,\,a]  +i S_{ff}^-[\chi_1,\,\lambda_1,\,a+B]+iS_{ff}^+[\chi_2,\,\lambda_2,\,a]}\\
&\hspace{5.65cm}\leftrightarrow\int \CD\vphi ~ e^{iS_{WF}[\vphi,\,B]}\equiv\CZ_{\text{WF}}[B]
\end{split}
\end{align}
holds as an equivalence at the conformal point.
Once more, we have imposed Neumann boundary conditions on the dynamical gauge field $a$ and Dirichlet boundary conditions on the scalar.
These results are summarized in Table \ref{tab:seed2_tab}.
We should recall the procedure that maps from \eq{bos1-2} to \eq{bos2-2} and make sure that it is consistent with our boundary picture.

In the bulk, this duality can be derived from the first seed duality by promoting the background spin$_c$ valued connection $A$ to a dynamical field, $a$, introducing an ordinary background $U(1)$ field $B$, and adding $-iS_{BF}[a,B]-iS_{CS}[B]$ to the action.
Looking first at the scalar side of this procedure and starting with \eq{wf-flux-4}, it becomes useful to define a new recipe for moving from the first seed duality to the second in the presence of a boundary by rewriting the BF term:
\begin{quote}
\textbf{New Promotion:} Promote $A$ to a dynamical field, $a$, introduce a new background field $B$, and add $iS_{CS}[a]-iS_{CS}[a+B]$ to the action.
\end{quote}
The Chern-Simons terms should be understood throughout the process in their microscopic descriptions with appropriate boundary conditions such that they give rise to chiral modes on the boundary to satisfy \eq{CH-mech}.
Once more, we have introduced the combination whose gravitational Chern-Simons terms cancel one another.
Note the old and new promotions are completely equivalent in the bulk where there are no surface terms from integration by parts or chiral modes to consider on the boundary.

Applying this procedure to \eq{wf-flux-4} gives
\begin{align}\label{eq:alt-seed-two}
\int \CD\vphi \CD a\CD b ~ e^{iS_{WF}[\vphi,b]+iS_{CS}[b+a]-iS_{CS}[a+B]}.
\end{align}
For brevity, we will leave the process of rewriting Chern-Simons terms as fermion and Pauli-Villars fields as implied moving forward.
When integrating out the dynamical fields, we find in the absence of holonomies, an assumption we will always make from now on, $0=b+a$, $0=a+B$, and thus $b=-a=B$.

With the methods we used in the first seed duality, it is straightforward to establish a duality between non-anomalous theories in the second.
After integrating out the dynamical fields, there are no ordinary or gravitational Chern-Simons terms left over for either mass deformation.
This is easiest to understand on the scalar side.
There are no Chern-Simons terms present regardless of the mass deformation, and hence, there are no edge movers required for the theory to be non-anomalous.
Since the scalar fields give rise to no chiral edge modes, we are consistent with \eq{CH-mech}.

Following our process for promotion for the free fermion gives
\begin{align}\label{eq:alt-seed-fermion}
\CZ_{\text{fermion+flux}}[B] = \int \CD\psi\CD a ~  e^{iS_f[\psi,\,a] +iS_{CS}[a]-iS_{CS}[a+B]}.
\end{align}
In the IR limit, integrating out the fermion gives
\begin{equation} \label{eq:alt-seed-CS}
S_{\text{fermion+flux}} =
-\frac{1}{2}\left(1-\text{sgn}(m_\psi)\right) (iS_{CS}[a]+i\Omega)-iS_{CS}[a+B]+iS_{CS}[a].
\end{equation}
For $m_{\psi}>0$, integrating out the fermion gives no Chern-Simons terms, and the equations of motion for $a$ imply $B=0$; leaving behind no Chern-Simons terms and the edge modes of the fiducial fermions exactly cancel.
For $m_{\psi}<0$, the first and last Chern-Simons terms and DWFs cancel and we are left with a $-iS_{CS}[a+B]-i\Omega$.
Here we can again find a theory completely determined by its framing anomaly and hence it can be replaced by $+i\Omega$.\footnote{This follows in an analogous manner to \eq{triv_1}. To see this, rewrite the dynamical spin$_c$ valued connection as the sum of a background spin$_c$ valued connection and a dynamical $U(1)$ connection $a=b+A$. Then, we can simply shift away the extra $B$ to recover the usual expression.}
Microscopically, this amounts to
\begin{equation}
\int \CD\chi~\CD\lambda~\CD a ~ e^{iS_{ff}^{\pm}[\chi,\lambda,a+B]}=1.
\end{equation}
This leaves behind no ordinary or gravitational Chern-Simons terms and hence no edge modes are left behind.
Thus, we find that after integrating out the dynamical degrees of freedom requiring the absence of anomalies for each of the Chern-Simons terms individually gives us a consistent theory.

Note that the fiducial fermion picture may not seem strictly necessary in this duality since there are no nonzero Chern-Simons terms from mass deformations and hence no edge movers are necessary to make the theory non-anomalous.
However, the fiducial fermions \emph{do} play an integral role in the above analysis since they cancel the would-be dynamical DWF, which cannot be eliminated without additional edge movers.

As with the first duality, imposing boundary conditions on the scalar requires a closer look at the global symmetry currents.
Choosing Neumann boundary conditions on the dynamical gauge field $a$ implies a constraint to field configurations which obey $(\partial_y a_i-\partial_i a_y)|_{y=0}=0$.
This also means the topological current parallel to the boundary vanishes, since $j^{i}_{\text{flux}}\propto\partial_y a_i-\partial_i a_y$.
Since this topological current should be identified with the particle number current on the scalar side of the duality, consistency requires $j^{i}_{\text{scalar}}|_{y=0}=0$.
Again, this can only be achieved by imposing Dirichlet boundary conditions on the scalar.

Lastly, one can easily check consistency of the above prescriptions by applying the promotions again to get back to the first seed duality.
The only subtlety is the sign of all the Chern-Simons terms in the promotion need to be flipped.
This means that our prescription is to promote $B$ to a dynamical field in \eq{bos2-4}, introduce a new background field $A$, add $+iS_{CS}[b+A]-iS_{CS}[A]$ to the action, and integrate out the dynamical fields.
Following this through, we are left with the appropriate chiral modes for the remaining Chern-Simons terms to satisfy \eq{CH-mech}.

\subsubsection*{Time-reversed dualities}

The time-reversed version of the seed dualities follow in a completely analogous manner.
Since the Chern-Simons terms are time-reversal odd, in order to satisfy \eq{CH-mech} we also need to swap the chiralities of the fermionic boundary terms.
Other than the minor consistency check required by the fermionic and Pauli-Villars boundary conditions, the time-reversed analogs of \eq{bos1-2} and \eq{bos2-2} are
\begin{align}
	\begin{split}\label{eq:bos1-2-tr}
\bar{\CZ}_{\text{WF+flux}}[A]\equiv& \int \CD\vphi ~\CD b~ e^{iS_{WF}[\vphi,\,b] -i S_{CS} [b+A] +i S_{CS}[A]} \\ & \hspace{5.5cm}\leftrightarrow\int \CD\psi~\CD\lambda~e^{i\bar{S}^+_{f}[\psi,\lambda,\,A]}\equiv\bar{\CZ}_{\text{f}}[A],
\end{split}
\end{align}
and
\begin{align}
	\begin{split}\label{eq:bos2-2-tr}
\bar{\CZ}_{\text{f+flux}}[B]\equiv&\int \CD\psi~\CD\lambda~\CD a ~e^{i \bar{S}^-_f[\psi,\lambda,\,a]  +i S_{CS}[a+B]-iS_{CS}[a]}\\
&\hspace{5.65cm}\leftrightarrow\int \CD\vphi ~ e^{iS_{WF}[\vphi,\,B]}\equiv\bar{\CZ}_{\text{WF}}[B].
\end{split}
\end{align}
As in the previous versions of the dualities, we can simply identify the correct number of boundary modes needed to ensure the absence of anomalies by looking at the sign and level of the Chern-Simons term directly.

\subsection{Particle-Vortex duality} \label{sec:PV-dual}
\subsubsection*{Scalar-Vortex duality}
Moving deeper into the web of dualities in \cite{Karch:2016sxi,Seiberg:2016gmd}, we will start with finding the influence of a boundary on
\begin{align}\label{eq:boson-vortex}
\overline{\CZ}_{\text{WF}}[C] \leftrightarrow \CZ_{\text{scalar-QED}}[C].
\end{align}
Beginning with \eq{bos2-4}, this duality is derived by promoting $B$ to be dynamical, introducing a new background field $C$, and adding $-iS_{CS}[b]+iS_{CS}[b+C]-iS_{CS}[C]$ to both sides of the duality.
Note these terms are equivalent to $iS_{BF}[b,C]$ in the absence of boundaries.
However, there would appear to be an issue of applying our fiducial fermion prescription to this duality.
That is we have Chern-Simons terms of ordinary $U(1)$ -- rather than spin$_c$ valued -- connections.\footnote{Recall, a $U(1)$ Chern-Simons term is well defined modulo $\pi\mathbb{Z}$ in general. It is only picking a spin structure that makes it well defined modulo $2\pi\mathbb{Z}$.}
The coupling of the fiducial fermions to such fields violates the relation forced by \eq{spin-charge} discussed in \sn{spin}.
However, we can work around that by rewriting the BF term including a spin$_c$ valued connection as \cite{Hsin:2016blu}
\begin{align}\label{eq:bf-cs2}
S_{BF}[b,C] = S_{CS}[b+C+A] - S_{CS}[b+A]- S_{CS}[C+A]+S_{CS}[A].
\end{align}
Note that all of the gravitational Chern-Simons terms that would have accompanied each $S_{CS}$ on the right hand side of \eq{bf-cs2} cancel and have thus been ignored.
Now, the promotion of the ordinary background connection, $B\rar b$, and the subsequent coupling to another ordinary background connection $C$ can be realized as a system of four fiducial fermions in the usual way.

Proceeding with the prescription, the scalar side of the duality becomes
\begin{align}\label{eq:particle-vortex-scalar-qed}
\CZ_{\text{scalar-QED}}[C] = \int \CD\vphi \CD b\, e^{iS_{WF}[\vphi,b]+iS_{CS}[b+C+A]-iS_{CS}[b+A]-iS_{CS}[C+A]+iS_{CS}[A]}.
\end{align}
The analysis of Chern-Simons terms and edge modes follows in a similar fashion to the WF $+$ flux case.
In the phase where the scalar is massive, the equations of motion for $b$ imply $C=0$, which causes the four Chern-Simons terms and associated edge modes cancel.
In the Higgsed phase, $b=0$, and once more all Chern-Simons terms cancel and there are no edge modes.

The modified fermionic theory is
\begin{align}\label{eq:particle-vortex-scalar-1}
\hspace{-0.2cm}\CZ_{\text{f}^{\hspace{0.025cm}\prime}}[C] =\int \CD\psi \CD\lambda \CD a\,\CD b\, e^{iS_{f}^{+}[\psi,\lambda,a] +iS_{CS}[a]-iS_{CS}[a+b]+iS_{CS}[b+C+A]-iS_{CS}[b+A]-iS_{CS}[C+A]+iS_{CS}[A]}.
\end{align}
Integrating out $b$ implies $b=C-a$ and plugging this back into the above expression yields
\begin{align}\label{eq:particle-vortex-scalar-2}
\CZ_{\text{f}^{\hspace{0.025cm}\prime}}[C] =\int \CD\psi\CD\lambda\,\CD a\, e^{iS_{f}^{+}[\psi,\lambda,a] +iS_{CS}[a-C]}.
\end{align}
Up to the sign of the mass terms, the two terms in the action of \eq{particle-vortex-scalar-2} are exactly the time-reversed alternate seed duality, \eq{bos2-2-tr}, with $B\to -C$, so that\footnote{The $-iS_{CS}[a]$ term is hidden in our difference of Pauli-Villars masses in $S_f$ and $\bar{S}_f$.}
\begin{align}
\CZ_{\text{f}^{\hspace{0.025cm}\prime}}[C] = \overline{\CZ}_{\text{f+flux}}[-C] \leftrightarrow \overline{\CZ}_{\text{WF}}[C].
\end{align}
This confirms the desired relation in \eq{boson-vortex}.
This is consistent with the scalar-QED side of the duality.

There is one caveat to the use of the time reversed duality connected to our use of $\overline{\CZ}_{\text{WF}}$ rather than $\CZ_{\text{WF}}$.
The time reversal operation changes the sign on the fermion mass term.
This has the effect of flipping the relationship between the way mass deformations in the two scalar theories are mapped to one another: positive mass deformations in $\overline{\CZ}_{\text{WF}}$ correspond to {\emph{negative}} mass deformations in $\CZ_{\text{scalar-QED}}$.
However, at the conformal fixed point $\overline{\CZ}_{\text{WF}}$ is completely equivalent to $\CZ_{\text{WF}}$.
This is a nice check, since it reproduces the equivalence $m_{\vphi}^2\leftrightarrow -m_{\vphi^{\prime}}^2$ on the two sides of the bosonic particle-vortex duality.

\subsubsection*{Fermion-Vortex duality}

The last duality we will consider in the presence of a boundary is the fermionic particle-vortex duality, which has some additional nuances.
This duality,
\begin{equation} \label{eq:ferm-pv}
\overline{\CZ}_{\text{f}}[A]e^{-\frac{i}{2}S_{CS}[A]}\leftrightarrow\CZ_{\text{QED}_3}[A],
\end{equation}
was originally formulated with theories which are $\mathcal{T}$-invariant on both sides, similar to the bosonic case \cite{Son:2015xqa}.

Recall that with our definition of $\CZ_{f}$ in \eq{dirac} this partition function contains the contribution
of the negative mass, heavy Pauli-Villars field $\lambda$.
Often the regulator is treated as producing a level -$\frac{1}{2}$ Chern-Simons term when integrated out.
More precisely, we get the $\eta$-invariant of $A$.
This factor means that $\CZ_{\text{f}}$ is not time reversal invariant: $m_\lambda \rar -m_\lambda$.
The purpose of the $e^{-\frac{i}{2} S_{CS}[A]}$ in \eq{ferm-pv} is to cancel the $\eta$-invariant and produce a time-reversal invariant fermionic partition function.
However, from our normalization in \eq{cs} we require that $k\in\mathbb{Z}$ for the Chern-Simons term to be gauge-invariant.
Thus, multiplying with half-integer Chern-Simons terms is not a consistent procedure in a purely $2+1$ dimensional theory.
To avoid this issue, this term can be viewed as arising as a boundary insertion in a theory on a $3+1$ dimensional bulk manifold, $X$ \cite{Metlitski:2015eka,Wang:2015qmt,Seiberg:2016gmd}.
More precisely, one promotes $A$ to a spin$_c$ valued connection on $X$ and adds
\begin{equation}
\frac{1}{8\pi}\int_X dA\wedge dA
\end{equation}
to the Lagrangian.
This promotion of $A$ to a spin$_c$ valued connection is possible for any (orientable) choice of bulk $X$ as all such $3+1$ dimensional manifolds admit a spin$_c$ structure.
This cancels the contribution of the regulator; rendering the fermionic partition function real and both sides of the duality time-reversal invariant.
All of this is perfectly valid in the $2+1$ dimensional bulk, but in the present context -- where $\Rplus$ would need to be realized as a boundary surface -- this prescription fails.
Indeed, had we proceeded through with multiplying $\CZ_{\text{f}}$ in with $e^{\frac{i}{2} S_{CS}[A]}$ as in \cite{Karch:2016sxi}, we would have found the Chern-Simons levels of $\pm\frac{1}{2}$ on either side of the mass deformation.
This is a clear contradiction with the assertion that the boundary is non-anomalous: We cannot generate ``half'' a DWF to satisfy \eq{CH-mech}.

Thus we find that in order to have a purely $2+1$ dimensional description of fermionic particle-vortex duality, we must either abandon time-reversal invariance at the conformal fixed point or find some other means of canceling the $\eta$-invariant of $A$.

Let us first explore what happens when we give up time reversal invariance.
It is no longer necessary to transfer the $k=\frac{1}{2}$ Chern-Simons term from one side of the duality to the other.
In this case, it will be convenient to begin our derivation with \eq{bos2-4}.
We then promote the background field to be dynamical, $B\to b$, and couple to a new background spin$_c$ valued connection $A$ via $-iS_{CS}[b+A]+iS_{CS}[A]$, the fermion+flux side is
\begin{equation}\label{eq:QED3_prime}
\CZ_{\text{QED}_3^{\prime}}[A] = \int \CD\psi~\CD\lambda~\CD a~\CD b\,e^{iS_f^+[\psi,\lambda,a] -iS_{CS}[a+b]+iS_{CS}[a]-iS_{CS}[b+A]+iS_{CS}[A]}.
\end{equation}
where the prime is being used to distinguish this from $\mathcal{T}$-invariant QED$_3$.
We proceed as usual in the IR limit and integrate out the dynamical fields $a$ and $b$.\footnote{More precisely, we must integrate out $a$ before $b$ to avoid imposing conditions which violate the spin-charge relation of our connections, i.e. imposing $2b=-a-A$ \cite{Seiberg:2016gmd}. The same condition prevents us from simplifying \eq{QED3_prime} by integrating out $b$.}
For $m_{\psi}>0$ we find no Chern-Simons terms, while for $m_{\psi}<0$ we find $iS_{CS}[A]+i\Omega$.
The fiducial fermion associated with $iS_{CS}[A]$ provides the necessary right-mover.

Meanwhile, the scalar side yields
\begin{equation}
\CZ_{\text{scalar}^{\prime}}[A] = \int \CD\vphi~\CD b ~ e^{iS_{\text{WF}}[\vphi,\,b] -iS_{CS}[b+A]+iS_{CS}[A]}.
\end{equation}
However, we recognize this as the time-reversed first seed duality, \eq{bos1-2-tr}. This ultimately gives
\begin{equation}
\CZ_{\text{QED}_3^{\prime}}[A]\leftrightarrow\overline{\CZ}_{\text{f}}[A].
\end{equation}
Again, we end up with level-$0$ and $1$ ordinary and gravitational Chern-Simons terms on either side of the mass deformation.
This time, the dynamical fermion can provide consistent chiral edge modes satisfying \eq{CH-mech}.

The other way to proceed is to insist on time-reversal invariance at the fixed point and doubly quantize the fields to avoid issues associated with half-integer Chern-Simons terms.
With this redefinition of our fields, cancelling the $\mathcal{T}$-violating $\eta$-invariant term can be achieved with a term which meets the quantization requirements of \eq{cs}.
However, taking $A=2A^{\prime}$ for some new spin$_c$ valued connection $A^{\prime}$ is in violation of the spin-charge relation, which would mean such an effective theory is not relevant to usual condensed matter systems \cite{Seiberg:2016gmd,Seiberg:2016rsg}.

Following similar steps to that above, we find
\begin{align}
\begin{split}
	\CZ_{\text{QED}_3^{\prime\prime}}[A]\equiv\int \CD\psi~ \CD\lambda~\CD a ~e^{iS_f^+[\psi,\lambda,2a]-2iS_{CS}[a]+2iS_{CS}[a+A]-2iS_{CS}[A]}&\\
	&\hspace{-4.0cm} \leftrightarrow \int \CD\psi \CD\lambda ~e^{i\bar{S}_f^+[\psi,\lambda,\,2A]}= \bar{\CZ}_{\text{f}}[2A].
\end{split}
\end{align}
It is straightforward to show edge movers are consistent with \eq{CH-mech} with an ordinary $U(1)$ connection fiducial fermion prescription, analogous to \eq{FF_expr},
\begin{equation}\label{eq:FF_spin}
e^{\pm i S_{CS}[B]} = \int \CD\chi\,\CD\lambda~ e^{i S_{ff}^{\pm} [\chi,\,\lambda,B]}.
\end{equation}
One needs to keep in mind the double gauge field coupling causes the edge modes to contribute double the anomalous current, but this is still compensated by the Chern-Simons current inflow.

\section{Lattice construction}\label{sec:lattice}

In this section, we will build on recent work that realized the Abelian dualities in \cite{Karch:2016sxi,Seiberg:2016gmd} using exact techniques.
We will consider the complex XY model on a Euclidean cubic lattice in $d=3$ as in \cite{Chen:2017lkr}.
We will introduce a boundary to this formalism in order to find the microscopic description of one of the dualities described in \sn{dualities}, the claim that scalars with flux are equivalent to a theory of fermions.

Our conventions for the lattice will be that the matter living at lattice sites are denoted by a subscript $n$ and the link variables are labeled by $n\mu$ designated to mean pointing from site $n$ in the direction $\hat{\mu}$.
A boundary will be implemented by simply truncating the lattice in the $y$-direction, rendering it semi-infinite.
We use the index $\beta$ for sites on the boundary.
Link variables transverse and parallel to the boundary will be denoted by $\beta y$ and $\beta i\in\{\beta t,\,\beta x\}$, respectively.

To realize the scalar $+$ flux theory, we start with the XY model for a complex scalar living at lattice site $n$, $\Phi_n \sim e^{i\theta_n}$ given in terms of a set of phase variables $\theta_n\in [0,\,2\pi)$ and background $U(1)$ gauge fields living on links $A_{n\mu}$ by
\begin{align}\label{eq:lattice-xy}
\CZ_{\text{XY}}[A] =\Big(\prod_n\int_{-\pi}^\pi\frac{d\theta_n}{2\pi}\Big)~\text{exp}\Big\{ \frac{1}{T}\sum_{n\mu}\cos(\theta_{n+\hat{\mu}}-\theta_n-A_{n\mu})\Big\}\equiv \int {\cal D}\theta~e^{-\frac{1}{T}H_{XY}[A]}.
\end{align}
To generate the necessary Chern-Simons term, we will employ the trick of coupling \eq{lattice-xy} to two-component Grassmann fields $\chi_n$ and $\bar{\chi}_n$.
This is equivalent to our fiducial fermion prescription in the continuum case.
The fermionic sector of the theory is given by
\begin{align}\label{eq:lattice-fermion}
\CZ_{\text{W}}[A] &=\prod_n \int d^2\bar{\chi}_n d^2\chi_n ~ e^{-H_W[A](M) - H_{\text{int}}(U)},
\end{align}
where the Wilson action $H_W$ and hopping-hopping interaction $H_{\text{int}}$ are
\begin{subequations}
	\begin{align}\label{eq:lattice-wilson}
	-H_W[A](M)&=\sum_{n\mu}\big(D_{n\mu}e^{-iA_{n\mu}}+D_{n\mu}^{*} e^{iA_{n\mu}}\big)+\sum_n (M-R)\bar{\chi}_n\chi_n,\\\label{eq:lattice-int}
	-H_{\text{int}}(U)&=U\sum_{n\mu} D_{n\mu}D^*_{n\mu}.
\end{align}
\end{subequations}
with $D_{n\mu}$ and $D_{n\mu}^*$ the fermionic forward and backward hopping terms, respectively
\begin{equation} \label{eq:ferm_hop}
D_{n\mu}\equiv\Big(\bar{\chi}_n\frac{\sigma^\mu + R}{2}\chi_{n+\hat{\mu}} \Big),\qquad D_{n\mu}^*\equiv \Big(\bar{\chi}_{n+\hat{\mu}}\frac{-\sigma^\mu + R}{2}\chi_{n}\Big).
\end{equation}
This particular form of $H_{\text{int}}$ is chosen in \cite{Chen:2017lkr} to reproduce the known continuum results.
Similar to the continuum theory, integrating out these Wilson fermions will produce the Chern-Simons term.
However, as a consequence of fermion doublers, the level of the resulting Chern-Simons theory is dependent on the relative magnitudes of  $M$ and the Wilson term, $R$, as well as the sign of $R$.
Compiling the above components of the theory and including the analog of the dynamical $U(1)$ gauge field present in the continuum theory, the scalar coupled to flux is
\begin{align} \label{eq:lattice-full}
\CZ[A] = \int {\cal D}a~ \CZ_{\text{XY}}[a]\,\CZ_{\text{W}}[A-a],\qquad \int{\cal D}a \equiv \prod_{n\mu}\int_{-\pi}^\pi\frac{da_{n\mu}}{2\pi}.
\end{align}

For the remainder of this section, we will assume $|R|=1$, which is motivated by reflection positivity.
Additionally, we assume we have chosen $T$, $U\lesssim 0$, and $M\lesssim 6$ in order to hit the IR critical point, as explained in \cite{Chen:2017lkr}.\footnote{We have chosen to define eqs. (\ref{eq:lattice-wilson}) and (\ref{eq:lattice-int}) such that it matches \cite{Golterman:1992ub, Jansen:1992tw, Shamir:1993zy, Kaplan:1992bt} and thus differs slightly from that of \cite{Chen:2017lkr}. To translate back, take $(M-3R)\to M$ and then $R\to -R$.}
That is, these values are tuned such the theory \eq{lattice-full} flows in the IR to
\begin{align} \label{eq:lattice-fermion}
\CZ_{\text{W}}[A] &=\prod_n \int d^2\bar{\chi}_n d^2\chi_n ~ e^{-H_W[A](M^\prime) - H_{\text{int}}(U^\prime)},
\end{align}
with $M^\prime=6$ and $U^\prime=0$.
\subsubsection*{Boundary conditions}

To study the effect of the presence of a boundary on \eq{lattice-full}, we need to understand how boundary conditions come about on the site and link variables.
We will start with the scalar fields, $\Phi_\beta$.
Ideally, we would have a direct analogy to the continuum case where either Neumann or Dirichlet boundary conditions are possible.
The former can be implemented by requiring the scalar hopping terms perpendicular to the boundary vanish.
However, due to our construction of scalar fields as having magnitude one, $\Phi_n \sim e^{i\theta_n}$, it is not actually possible to enforce Dirichlet boundary conditions, i.e. $\Phi_{\beta}=0$.
Instead, we will enforce Dirichlet boundary conditions by requiring the scalar current along the boundary to be zero.

The fermionic boundary conditions are such that either
\begin{align} \label{eq:lattice-bc-fermion}
P_+ \chi_\beta = \bar{\chi}_\beta P_-=0,\qquad \text{or}\qquad P_-\chi_\beta = \bar{\chi}_\beta P_+ =0,
\end{align}
extremize the boundary variation term \cite{Sint:1993un}.
We will use as our convention $\sigma^{\hat{y}} = {\scriptsize{\begin{pmatrix} 1 & 0\\ 0 & -1\end{pmatrix}}}$ such that the chiral projectors in \eq{lattice-bc-fermion} are $P_\pm = \frac{1}{2}(1\pm\sigma^{\hat{y}})$.
From the assumption that $|R|=1$ and up to a sign, the chiral projectors are equivalent to the matrices $\frac{1}{2}(\pm \sigma^{\hat{y}} - R)$ appearing in the fermionic hopping terms perpendicular to the boundary in eqs. (\ref{eq:lattice-wilson}) and (\ref{eq:lattice-int}).
Either of the conditions in \eq{lattice-bc-fermion} will remove one chiral mode worth of degrees of freedom, while the other chiral mode is left unconstrained.
These conditions can be compared to those in \eq{ferm-bc} and be seen to agree -- albeit by construction \cite{Sint:1993un}.

Lastly, we need to consider the link variables.
We again draw inspiration for the appropriate boundary conditions from the continuum case.
That is, Neumann boundary conditions correspond to the condition that plaquettes perpendicular and adjacent to the boundary must vanish.
On the lattice, this will correspond to the constraint
\begin{align}\label{eq:lat-gauge-bc}
a_{\beta i}+a_{(\beta+\hat{i})y}-a_{(\beta+\hat{y})i}-a_{\beta y}=0.
\end{align}
Alternatively, we could choose Dirichlet boundary conditions which simply require $a_{\beta i}=0$.
%
We would like to reproduce the results of the continuum duality \eq{bos1-4} and this will guide us in choosing the corresponding boundary conditions on the lattice.

\subsubsection*{Implementation}

The main results of ref. \cite{Chen:2017lkr} -- following the choice of hopping-hopping interaction $H_{\text{int}}$ -- are contained in the identification of a suitable UV map for the conserved currents built out of $\theta_n$ and $a_{n\mu}$ into a theory of free fermions.
Those theories are then flowed to the IR where one can then compare to continuum results.
Following these general principles, we identify the effects of truncating the lattice at some arbitrary boundary site.
We will show the derivation of \cite{Chen:2017lkr} holds in the presence of a truncated boundary and is non-anomalous for $M>0$ so long as $R=1$ and the $P_-\chi_\beta=\bar{\chi}_\beta P_+=0$ boundary condition is chosen.
We will also verify mass deformations away from the conformal fixed point yield equivalent results to the continuum case.

Recall the existence of a DWF at the boundary was of particular importance in our continuum picture for self-consistency checks away from the conformal point.
A truncated lattice also gives rise to massless chiral modes localized to the boundary \cite{Shamir:1993zy}.
In particular, there are fermionic modes obeying
\begin{equation}
\Psi_{\pm}(x,y,t)=\xi(y)(1\pm\sigma^{\hat{y}})\psi_{\pm},\qquad \xi(y)\equiv[1-F^2(k_\mu)]^{\frac{1}{2}}[F(k_\mu)]^y
\end{equation}
with $\psi_{\pm}$ a right/left helicity eigenstate and
\begin{equation} \label{eq:dwf_lat}
F(k_i)=R-M+R\sum_{i=t,x}(1-\cos k_i).
\end{equation}
For a given $k_i$, this solution can be normalized only if $|F(k_i)|<1$ \cite{Shamir:1993zy}. At the limit $|F(k_i)|=1$ the DWF becomes a continuum eigenstate.

Now let's turn to the derivation of the duality.
We will follow the derivation of ref. \cite{Chen:2017lkr}  and point out where subtleties of the boundary come into play.
To begin, rewrite the bosonic hopping term to make the bosonic currents explicit
\begin{align} \label{eq:lattice-boson-current}
e^{\frac{1}{T}\cos(\theta_{\beta+\hat{\mu}} - \theta_{\beta}-a_{\beta\mu})} = \sum_{j_{\beta\mu}=-\infty}^\infty I_{\beta_\mu}(T^{-1})\Phi_\beta\Phi_{\beta+\hat{\mu}}^*e^{-ia_{\beta\mu}j_{\beta\mu}},
\end{align}
where $I_j$ is the $j^{\tiny{\text{th}}}$ modified Bessel function.
As mentioned in the previous section, we will enforce Dirichlet boundary conditions on the scalar by requiring the scalar current in the boundary to vanish, i.e. $j_{\beta i}=0$.
The bosonic degrees of freedom can be integrated out explicitly and this simply enforces Gauss's law for the scalar currents at the boundary sites.
By current conservation, this implies current onto the boundary also vanishes, $j_{\beta y}=0$.

The implementation of boundary conditions for the Grassmann variable and their effect on eqs. (\ref{eq:lattice-wilson}) and (\ref{eq:lattice-int}) is more subtle.
In the continuum case, one of these boundary conditions will kill off the DWF on the boundary, while the other will leave it untouched.
This had important implications relating to the anomalous nature of the theory.
Is this feature also realized in the lattice?
To see this is still consistent with the Callan-Harvey mechanism on either side of the mass deformation, we need to take a closer look at the interplay between Chern-Simons terms and DWFs on the semi-infinite lattice.

On the lattice, the Chern-Simons term is determined by the masses of the $2^3=8$ chiral Dirac fermion modes in the continuum.
These correspond to the eight extrema of the Brillouin zone at $k_{t,x,y}=\lbrace0,\pi\rbrace$.
The effective masses of these eights modes are determined by \cite{Golterman:1992ub,Chen:2017lkr}
\begin{equation} \label{eq:lat_meff}
m_{\text{eff}}(k_\mu)=M-R\sum_\mu(1-\cos k_\mu).
\end{equation}

Since the value of $R$ is important in eqs. (\ref{eq:dwf_lat}) and (\ref{eq:lat_meff}), we should see if we can first fix its sign.
Recall that it is the current of the Chern-Simons term flowing onto the boundary which renders the theory non-anomalous.
This current is nonzero only when the $R$ and $M$ in \eq{lattice-int} have the same sign \cite{Golterman:1992ub}.
Hence, given our choice of $M>0$ we must take $R=1$ to allow for anomaly inflow.

From \eq{ferm_hop}, the choice of $R=1$ has the effect of projecting onto the right-moving chiral mode for hopping terms perpendicular to the boundary.
For reasons that will become clear shortly, the correct fermionic boundary condition to choose in this case is $P_-\chi_\beta=\bar{\chi}_\beta P_+=0$.
Together with the choice of $R$, this implies $D^*_{\beta y}\ne0$ and $D_{\beta y}\ne0$ in general.
Had we chosen the opposite boundary conditions or $R=-1$ we would have found no current flow onto the boundary.

With $R$ fixed, the value of $M$ -- or equivalently, $M^{\prime}$ of \eq{lattice-fermion} -- determines both the Chern-Simons level and the existence of DWFs for each of the $k_\mu$.
For our present purposes, we will only be concerned with the behavior of the theory in the vicinity of the critical mass, $M=6$, and so we will check the behavior of the $k_{\mu}$ extrema for these values.

\begin{table}
\begin{centering}
\begin{tabular}{ccccc}
\hline
\textbf{Brillouin Zone} & \textbf{Chirality} & \multicolumn{3}{c}{\textbf{Mass Parameter}}\tabularnewline
 \textbf{Extremum},$\,k_{\mu}$ &  & $M=6-\epsilon$ & $M=6$ & $M=6+\epsilon$\tabularnewline
\hline
$\left(0,0,0\right)$ & $R$ & $+$ & $+$ & $+$\tabularnewline
$\left(0,0,\pi\right)$ & $L$ & $-$ & $-$ & $-$\tabularnewline
$\left(0,\pi,0\right)$ & $L$ & $-$ & $-$ & $-$\tabularnewline
$\left(\pi,0,0\right)$ & $L$ & $-$ & $-$ & $-$\tabularnewline
$\left(0,\pi,\pi\right)$ & $R$ & $+$ & $+$ & $+$\tabularnewline
$\left(\pi,\pi,0\right)$ & $R$ & $+^{*}$ & $+$ & $+$\tabularnewline
$\left(\pi,0,\pi\right)$ & $R$ & $+$ & $+$ & $+$\tabularnewline
$\left(\pi,\pi,\pi\right)$ & $L$ & $+^{*}$ & $0$ & $-$\tabularnewline[0.25cm]
\hline
\textbf{Total CS Level} &  & $1$ & $\frac{1}{2}$ & $0$\tabularnewline
\textbf{Total DWF} &  & one $R$, one $L$ & none & none\tabularnewline
\hline
\end{tabular}
\par\end{centering}
\caption{Chirality, mass, and existence of a DWF for the eight modes at the extremum of the Brillouin zone, $k_{\mu}=(k_t,k_x,k_y)$, as calculated using eqs. (\ref{eq:dwf_lat}) and (\ref{eq:lat_meff}). Positive and negative masses are denoted by a $+$ and $-$, respectively and an astrix denotes a mode which meets the condition to be a DWF.\label{tab:dwf_cs}}
\end{table}

Our results are summarized in Table \ref{tab:dwf_cs}.
For $M=6$, corresponding to the IR fixed point, the Chern-Simons term is level-$\frac{1}{2}$ and there are no DWFs.
More precisely, the would-be DWF is at the limit where $|F(k_i)|=1$ and has become a continuum eigenstate.
This is consistent with the proposed continuum behavior at the conformal fixed point.
For $M^\prime=6+\epsilon$, the Chern-Simons level is zero and we have no DWFs since \eq{dwf_lat} is not satisfied for any $k_i$.
Again, this is in agreement with \eq{CH-mech}.

$M^\prime=6-\epsilon$ is slightly more subtle.
This value corresponds to the UV sector of the theory where we need to level-$1$ Chern-Simons to generate the $e^{iS_{CS}[A-a]}$ term as well as negative mass deformations at the IR fixed point.
For this case we find a Chern-Simons level of 1 and two DWFs, since both $k_{\mu}=(\pi,\pi,0)$ and $k_{\mu}=(\pi,\pi,\pi)$ satisfy \eq{dwf_lat}.
However, this is where our fermionic boundary conditions we enforced earlier come back into play.
Since we have a Chern-Simons level of $1$, we have chosen our boundary condition to kill off the left-mover, namely $P_-\chi_\beta=\bar{\chi}_\beta P_+=0$.
This gives the correct chiral modes on the boundary to satisfy \eq{CH-mech}.
Interestingly, since they supply a level-$\frac{1}{2}$ Chern-Simons term with no DWF, it is the fermion doublers that play the role of the Pauli-Villars regulator on the lattice.
Thus, we are self-consistent with the Callan-Harvey mechanism all the way through.
This analysis follows similarly for $M<0$ in which case we would need to choose $R=-1$ and kill off right-movers with the fermionic boundary condition.

Note that by imposing the fermionic boundary conditions, we have fixed two of the Grassmann variables we would normally integrate over on the boundary sites.
The fermionic current conservation imposed by Grassmann integration will still hold for such links, but now each site has only two Grassmann degrees of freedom instead of four.
The contribution of the double hopping/interaction term with any boundary site is very limited in such cases, since it already contains both Grassmann degrees of freedom.
To have a non-vanishing contribution it must be isolated from any other links.

Finally, we need to understand the effect of the Neumann boundary conditions on the dynamical gauge field, i.e. \eq{lat-gauge-bc}.
The bulk integration over the link variables tied the bosonic and fermionic currents together.
From the above construction, the boundary scalar current vanishes, which would seem to imply the boundary fermionic current does as well.
This would present a problem for satisfying \eq{CH-mech} if not for the gauge field boundary conditions.
Enforcing \eq{lat-gauge-bc} on, e.g., the $\beta i$ link kills off the link integration along the boundary and transforms the fermionic current terms as
\begin{align}\label{eq:lat-ferm-j}
D_{\beta i} e^{-i(A_{\beta i}-a_{\beta i})} \qquad\to\qquad D_{\beta i} e^{-i(A_{\beta i}+a_{(\beta+\hat{i})y}-a_{(\beta+\hat{y})i}-a_{\beta y})}.
\end{align}
Hence, there is no tying of the fermionic current to the vanishing scalar current, but we should still verify that is possible to get a non-vanishing fermionic current on the boundary so our DWFs are still allowed solutions.

First, consider $e^{i a_{\beta y}}$, which na\"ively would be problematic for the survival of terms like \eq{lat-ferm-j} upon integration over the corresponding link variable unless it is canceled by $e^{-i a_{\beta y}}$ from somewhere else in the path integral.
With no scalar current flowing onto the boundary, we could use fermionic current term such as $D^*_{\beta y} e^{i(A_{\beta y}-a_{\beta y})}$ to cancel $e^{i a_{\beta y}}$.
However, such a term means the fermionic current flows off the boundary.
Since the number of Grassmann variables at the site $\beta$ is saturated by the two fermionic currents due to our fermion boundary conditions,  a double-hopping term to return the fermionic current to the same site is forbidden.
Relying on such a cancellation would mean the boundary fermionic current is only supported for a single link.

Fortunately, there are additional contributions that work to cancel $e^{i a_{\beta y}}$.
Consider the form of \eq{lat-ferm-j} for neighboring boundary links.
The $(\beta-\hat{i})i$ link contains an exponential of the form $e^{-i a_{\beta y}}$ which can cancel $e^{i a_{\beta y}}$.
This has the interpretation of a fermionic current flowing from the $(\beta-\hat{i})i$ link to the $\beta i$ link.
The cancellation generalizes over a chain of adjacent boundary links with nonzero fermionic current and causes all exponentials with dynamical gauge links perpendicular to the boundary to vanish.

The only remaining term the needs to be cancelled in \eq{lat-ferm-j} is $e^{-ia_{(\beta+\hat{y})i}}$.
This can easily be achieved by either the fermionic or bosonic currents living on the $(\beta+\hat{y})i$ link.
Combining this with the cancellation of $e^{i a_{\beta y}}$ and $e^{-ia_{(\beta+\hat{i})y}}$, it is possible to have an uninterrupted fermionic current flowing along the boundary in spite of having chosen scalar boundary conditions which set bosonic currents on the boundary to zero.\footnote{It is also possible to have a nonzero fermionic current on the boundary with Dirichlet boundary conditions on the gauge field. This is still consistent with the continuum case, but would require killing off edge movers in order to get a non-anomalous theory.}
Furthermore, the chirality of this boundary current is set by our choice of fermionic boundary conditions.
This is completely analogous to the continuum case.

\section{Discussion and Conclusion}\label{sec:discuss}

In this work, we presented a generalization of Abelian bosonization that remains valid in the presence of a boundary.
Our main finding is that, for the duality to be valid in the presence of boundaries, one carefully needs to account for edge modes that are associated with Chern-Simons terms.
Most importantly, we require edge modes even for Chern-Simons terms in the action that only involve non-dynamical fields.
We implemented this consistently by replacing all Chern-Simons terms with heavy ``fiducial" fermions.

Given the fact that even the $S_{BF}[b,C]$ term of \eq{bf-cs2} can be rewritten using our fiducial fermion prescription to yield a consistent theory, a natural question one might ask if this is always the case.
In other words, can we ever run into some combination of Chern-Simons terms which is consistent with the spin-charge relation of a spin$_c$ but cannot be rewritten in terms of our fiducial building blocks?
Reassuringly, the answer appears to be no.
In \cite{Hsin:2016blu} it was shown that any consistently quantized Chern-Simons term which can be put on a spin$_c$ manifold can be rewritten as
\begin{subequations}
	\begin{align}\label{eq:spin_c-rewrite}
	S_{BF}[B,C] &= S_{CS}[B+C+A] - S_{CS}[B+A]- S_{CS}[C+A]+S_{CS}[A],\\\
	S_{CS}[B]+S_{BF}[B,A]& = S_{CS}[A+B]-S_{CS}[A],\\\
	16\text{CS}_{\text{g}} & = 9S_{CS}[A]-S_{CS}[3A].
	\end{align}
\end{subequations}
All such terms lend themselves to a description in terms of fiducial fermions.

From a condensed matter perspective, the fermionic particle/vortex was originally proposed as a $\CT$-symmetric UV completion of the half-filled lowest Landau level.
However, the need to view it as the surface of a $3+1$ dimensional topological insulator lead the authors of \cite{Wang:2015qmt,Wang:2015hfll} to conclude that there is no strictly $2+1$ dimensional UV completion for this system.
Our analysis suggests such a completion does exist so long as one is willing to lose the spin-charge relation of a spin$_c$ valued connection or $\CT$-invariance.
One can ask whether the projection onto the lowest Landau level is somehow inconsistent with formulating the theory on a spin$_c$ manifold.
If such inconsistencies arise, then the doubly quantized theory would provide a purely $2+1$ dimensional UV completion that is manifestly $\CT$ invariant.
This would require a rigorous study of lowest Landau level projectors on spin$_c$ manifolds -- a problem we leave to future work.

Since there have been other microscopic descriptions of the bulk Abelian dualities, e.g.~\cite{Mross:2017gny,Mross:2015idy}, one could wonder how those models realize the boundary physics as presented above.
In \cite{Mross:2017gny}, a discrete 2+1 dimensional lamination of 1-dimensional quantum wires was used to derive the Abelian bosonization and particle-vortex duality.
Each wire supporting a 1+1-dimensional continuum theory suggests a natural microscopic realization of the above results; the study of which is also left for future work.

Obvious questions we have not addressed in this work are generalizations to the non-Abelian case or to theories with interfaces rather than boundaries. We anticipate that they work in a similar way, but of course they come with extra subtleties that will be important to understand.  Lastly, left unexplored in this analysis among the transitions enumerated in \cite{Liendo:2012hy} are the ``extraordinary'' type where the boundary scalar gets a vev and drives a surface transition in addition to gapping the bulk.  That the extraordinary transition is believed to admit no relevant boundary deformations sets it apart from the boundary conditions studied in this work and warrants further study in the context of the $2+1$ dimensional dualities studied here. A rich network of dualities making along these lines has been laid out in \cite{gaiotto} based on conjectures about the infrared behavior of ``duality walls". It would be very interesting to generalize our work to these other options as well.

\section*{Acknowledgments}

We would like to acknowledge Anton Andreev, David Kaplan, Steve Sharpe, David Simmons-Duffin, Michael Wagman, and Larry Yaffe for helpful conversations as well as Kristan Jensen, David Mross, Ami Katz, and Nathan Seiberg for helpful email correspondence. Special thanks to Davide Gaiotto for pointing out important aspects of his work presented in \cite{gaiotto} and, most importantly, for emphasizing to us the importance of global symmetries for determining the correct boundary conditions for scalars and gauge fields. This work was supported, in part, by the U.S.~Department of Energy under Grant No.~DE-SC0011637.  In addition, the work of BR was funded, in part, by STFC consolidated grant ST/L000296/1.

\bibliographystyle{JHEP}
\bibliography{bdry}
\end{document}